# CI-Design #1: Silence

*Silence Routing: When Not Speaking Improves Collective Judgment*

*Itsuki Fujisaki[1], Kunhao Yang[2]

[1]College of Knowledge and Library Science, School of Informatics, Tsukuba University, Ibaraki, Japan
[2]College of Engineering, Shibaura Institute of Technology; Tokyo, Japan
*Corresponding authors: Itsuki Fujisaki
Email: bpmx3ngj@gmail.com +81-80-3022-6288 (I.F)

**Abstract**

The wisdom of crowds has been shown to operate not only for factual judgments but also in matters of taste, where accuracy is defined relative to an individual's preferences. However, it remains unclear how different types of social signals should be selectively used in such domains. Focusing on a music-preference dataset in which contributors provide both personal evaluations (Own) and estimates of population-level preferences (Estimated), we propose a routing framework for collective intelligence in taste. The framework specifies when contributors should speak, what they should report, and when silence is preferable. Using simulation-based aggregation, we show that prediction accuracy improves over an ALL-OWN baseline across a broad region of the parameter space, conditional on items where routing applies. Importantly, these gains arise only when silence is allowed, enabling second-order signals to function effectively. The results demonstrate that collective intelligence in matters of taste depends on principled signal routing rather than simple averaging.

## 1. Introduction

Classic research on the wisdom of crowds demonstrates that aggregating judgments across individuals can produce remarkably accurate collective outcomes (Galton, 1907; Surowiecki, 2004; Hertwig, 2012). More recent work has established that this phenomenon also extends to matters of taste, such as music or aesthetic preferences, where the goal is not to recover an objective truth but to predict an individual's subjective evaluation (Müller-Trede et al., 2018; Analytis et al., 2018). These findings suggest that aggregation can be useful even when correctness is personalized.

At the same time, it is well known that using all available judgments indiscriminately is not always optimal. Selection-based approaches show that carefully chosen subsets of individuals can outperform larger crowds (Mannes et al., 2014; Goldstein et al., 2014), and that moderately sized groups may be superior when task difficulty varies (Galesic et al., 2018). Related work on decision similarity further demonstrates that, even without feedback, individuals who agree more with others tend to be more accurate in binary decision tasks, allowing high-performing groups to be identified via similarity alone (Kurvers et al., 2019).

However, Kurvers et al. (2019) also document important boundary conditions. In particular, similarity-based selection can break down under structured correlations and in so-called wicked regimes where the minority is correct. These limitations raise a broader question: how should disagreement be treated when selection becomes brittle?

Rather than viewing disagreement purely as noise to be discarded, recent work has begun to explore whether dissimilar judgments can sometimes be transformed into informative signals. In estimation and binary-choice settings, Fujisaki and Yang (2026a) propose an inversion-based aggregation strategy that flips strongly dissimilar judgments before aggregation, showing that transformation and selection constitute distinct design levers. In matters of taste, Fujisaki and Yang (2026b) further demonstrate that dissimilar preferences can be conditionally informative: particularly for non-mainstream users when handled through appropriate transformations rather than exclusion.

A complementary literature emphasizes second-order information, such as beliefs about what others think. Meta-prediction methods, including the surprisingly popular approach and its Bayesian extensions, show that eliciting population-level expectations can correct shared biases and extract latent structure from the crowd (Prelec et al., 2017; McCoy & Prelec, 2024; Wilkening et al., 2022; Zhang et al., 2024). Performance-weighted and expertise-identification frameworks further highlight the importance of how information is used, not just who provides it (Himmelstein et al., 2023; Atanasov & Himmelstein, 2023; Collins et al., 2023).

Despite these advances, the possibility that contributors might offer different kinds of signals—or choose not to contribute at all—has received little systematic attention, especially in taste domains. In matters of taste, individuals can plausibly report their own preference, their belief about population-level taste, or remain silent when uncertain. Prior studies on taste discrimination show that people vary in how clearly they can evaluate a stimulus, and that this clarity shapes the usefulness of their judgments (Yaniv et al., 2011). Yet existing approaches do not address how such heterogeneity should guide the routing of signals.

In this paper, we propose a routing perspective on collective intelligence in matters of taste. Rather than treating judgments as homogeneous inputs, we explicitly model three options—Own, Estimated, and Silence—and ask how they should be combined. Using a music-preference dataset in which both Own and Estimated judgments are available for the same items (Fujisaki et al., 2022), we show that principled routing can systematically improve prediction accuracy, whereas strategies that ignore silence fail.

## 2. Method Overview

We reanalyze the music-preference dataset reported by Fujisaki et al. (2022), consisting of 56 participants who evaluated 24 musical excerpts. Each excerpt was truncated to approximately one minute, and the stimulus set spans diverse genres, including metal, pop, electronic, and traditional music. For each excerpt, participants reported (i) their own preference (Own), (ii) their estimate of how much people in general would like the excerpt (Estimated), and (iii) how difficult it was to determine their own preference. Participants responded to these questions on a scale ranging from 0–100.

A key intuition behind our routing framework is that disagreement between Own and Estimated judgments can be informative rather than noisy. When a participant finds an item easy to judge (e.g., difficulty = 15) but observes that their Own evaluation (e.g., 80) diverges strongly from their Estimated judgment (e.g., 20), this may indicate awareness that their personal taste is atypical. In such cases, reporting Estimated rather than Own may provide a more reliable signal for predicting another person's preferences. Conversely, when an item is difficult to judge (e.g., 90), contributors may lack a stable internal preference, and forcing them to provide either Own or Estimated judgments is likely to inject noise. Allowing silence in these cases can therefore protect the aggregate.

Operationally, routing proceeds in two stages. First, a difficulty-based participation rule determines eligibility: for a threshold $q$, each participant contributes only on items they personally rank among their easiest $q$ proportion; otherwise, they remain silent. Difficulty ratings are normalized within each participant to account for individual scale-use differences. Second, among eligible contributors, those with the largest divergence between Own and Estimated judgments are switched to provide Estimated judgments, while the remainder provide Own judgments. The parameter $K$ specifies the maximum number of contributors who may switch; for any given item, the effective number of Estimated signals is $min(K,$ number of eligible contributors$)$. Prediction accuracy is evaluated via mean squared error relative to an ALL-OWN baseline.

## 3. Results

Figure 1 maps changes in mean squared error (MSE) relative to the ALL-OWN baseline as a function of difficulty filtering strength q and the upper bound K on contributors switched to Estimated judgments. Each cell reports performance conditional on items for which at least one contributor is eligible under the routing rule.

Across a broad contiguous region of the parameter space, routing yields lower MSE than ALL-OWN. Improvements are not confined to a finely tuned optimum but arise whenever difficulty filtering excludes unreliable judgments and enough Estimated judgments are aggregated. In contrast, regions with weak filtering or very few Estimated contributors show little improvement or degradation. These patterns indicate that Estimated judgments are effective only when used collectively and under favorable reliability conditions.

Figure 2 provides a critical control by prohibiting silence. In this switch-only condition, contributors must always provide either Own or Estimated judgments. Under this constraint, no parameter setting produces stable improvements over ALL-OWN. This contrast demonstrates that silence is a necessary structural component of effective routing, rather than a minor refinement.

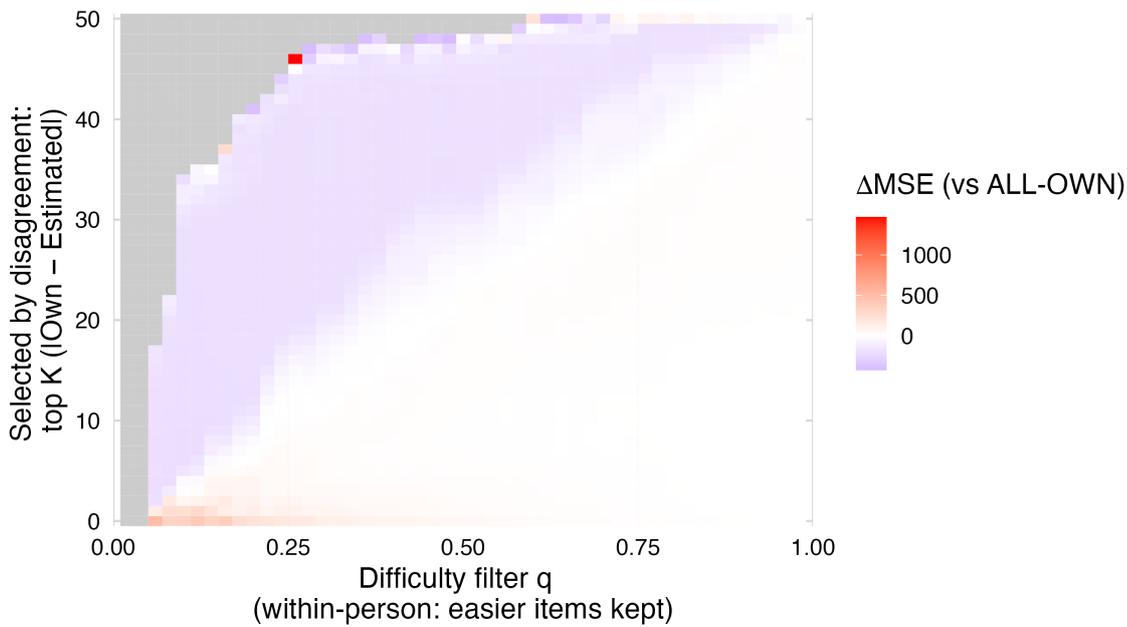

**Figure 2. Routing with Silence Improves Collective Accuracy in Matters of Taste.** Heatmap showing changes in mean squared error (ΔMSE) relative to the ALL-OWN baseline as a function of two routing parameters: the strength of difficulty-based filtering (q, x-axis) and the upper bound K on the number of contributors whose signals are switched from Own to Estimated (y-axis). Difficulty filtering is applied within each contributor, such that for a given q, a contributor provides input only for the proportion q of items they personally judge as easiest; otherwise, they remain silent. Among eligible contributors for each item, those with the largest divergence between Own and Estimated judgments are switched to Estimated, up to a maximum of K contributors. Importantly, K represents an upper bound: the effective number of Estimated signals per item is min(K, number of eligible contributors). Each cell reports ΔMSE conditional on items for which at least one contributor is eligible under the routing rule. Blue regions indicate improved accuracy relative to ALL-OWN. Improvements emerge across a broad contiguous region of the parameter space, indicating that routing enhances collective accuracy robustly rather than at a narrowly tuned optimum.

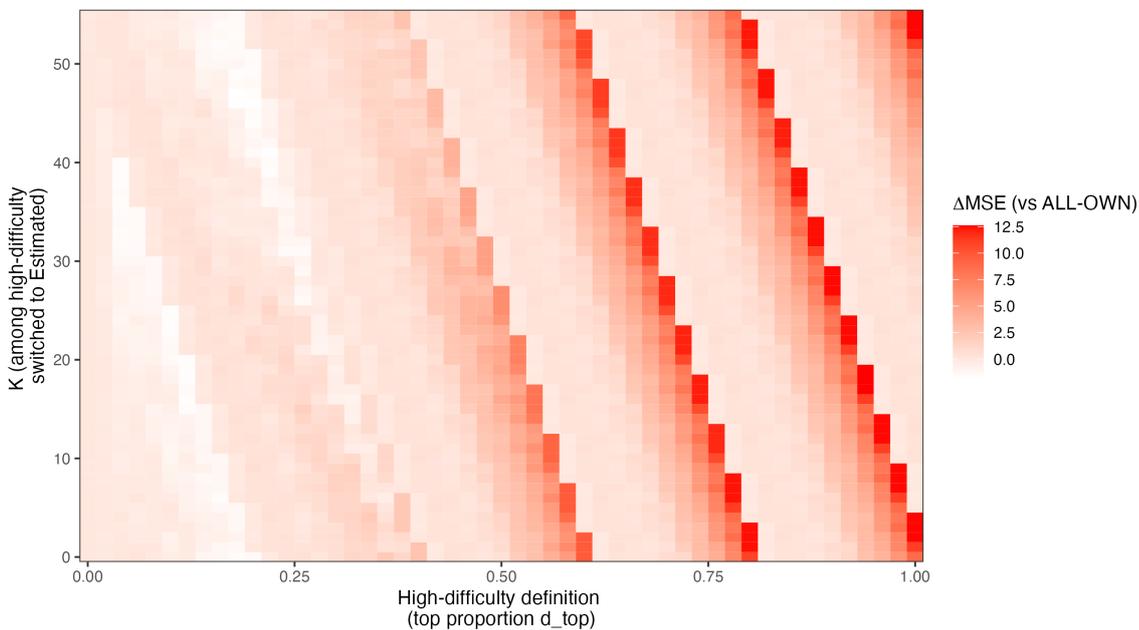

**Figure 1. Silence Is Necessary for Effective Signal Routing.** Control analysis corresponding to Figure 1, in which silence is prohibited. Contributors are forced to provide either Own or Estimated judgments for every item, following the same disagreement-based switching rule. Under this switch-only regime, no parameter setting yields stable improvements over the ALL-OWN baseline. This result demonstrates that allowing contributors to remain silent on difficult items is a necessary structural condition for second-order signals (Estimated judgments) to contribute constructively to collective prediction in matters of taste.

## 4. Discussion

Our findings extend research on collective intelligence in taste by highlighting routing as a central design principle. Prior work has shown that collective accuracy depends on who participates (Mannes et al., 2014; Himmelstein et al., 2023) and how information is weighted (Collins et al., 2023). We add that how signals are routed—and whether silence is permitted—can be equally consequential.

The role of meta-prediction is particularly nuanced. In objective-truth domains, population-level expectations can correct biases and reveal latent structure (Prelec et al., 2017; McCoy & Prelec, 2024; Wilkening et al., 2022; Zhang et al., 2024). In taste domains, however, Estimated judgments are not universally superior to Own judgments. Instead, they become useful only when aggregated from many contributors who are confident in their evaluations. This conditional usefulness explains why forcing Estimated judgments without silence fails.

Silence emerges as an enabling mechanism. Recent work shows that opt-out and abstention can improve group outcomes by limiting unreliable inputs (Kuroda et al., 2023; Mayer et al., 2025). Our results demonstrate that silence is essential for allowing second-order signals to function effectively in taste prediction.

The routing perspective also has cross-domain implications. In recommender systems, missing data are often treated as a problem to be imputed (Resnick et al., 1994; Resnick & Varian, 1997), which can amplify popularity and mainstream biases (Abdollahpouri et al., 2020; Analytis & Hager, 2023). Treating silence as informative and selectively aggregating population-level estimates may offer a complementary design principle.

## 5. Future directions

The routing framework introduced here opens several avenues for future research, both empirical and theoretical.

A first natural extension concerns diagnosing how routing operates at the item level. In the present analyses, performance improvements were evaluated conditionally on items for which routing applies, but the internal mechanics of this conditioning remain largely implicit. Visualizing, for each item, how many contributors are eligible to speak under different difficulty thresholds, and how many ultimately provide Own versus Estimated judgments, would make the routing process more transparent. Such analyses could clarify how improvements arise from heterogeneous participation patterns rather than uniform crowd behavior, and could be particularly informative for instructional or design purposes.

Second, while we evaluated accuracy primarily using mean squared error, other performance metrics may capture complementary aspects of collective prediction. For example, mean absolute error or rank-based measures may be less sensitive to outliers, while calibration-oriented metrics could reveal whether routing improves not only point accuracy but also the reliability of predictions. Exploring multiple error criteria would help establish the robustness of routing effects across evaluative lenses.

Third, the present work fixes the potential contributor pool but varies routing rules within that pool. Varying group size more systematically: by sampling different numbers of contributors, would connect routing to prior work on optimal crowd size and composition (e.g., Mannes et al., 2014; Galesic et al., 2018). Importantly, routing suggests that "group size" need not be a static parameter: effective group size can emerge endogenously through silence, with different items drawing on different subsets of contributors.

Fourth, routing invites deeper theoretical analysis of error structure. Decomposing prediction error into bias and variance components could clarify whether routing primarily reduces systematic bias, attenuates noise, or trades one for the other under different conditions (Müller-Trede et al., 2018; Fujisaki et al., 2022). Such decompositions may also help formalize why silence is necessary for second-order signals to become beneficial, and why switching alone fails.

Finally, a particularly promising direction concerns extension to numerical estimation tasks. In estimation domains, contributors can often provide not only point estimates but also confidence judgments and estimates of what others would report. While meta-prediction algorithms such as the surprisingly popular method have demonstrated success in some estimation settings (Prelec et al., 2017), their applicability to continuous estimation remains limited. The routing perspective suggests an alternative design logic: when a contributor's own estimate diverges strongly from their estimate of the population mean, confidence may determine whether that divergence reflects private information or idiosyncratic noise. High-confidence divergence may justify privileging the contributor's own estimate, whereas low-confidence divergence may call for averaging, down-weighting, or silence. Integrating confidence, self-estimates, population estimates, and silence within a unified routing framework could thus bridge individual-level and collective-level wisdom-of-crowds effects across both estimation and taste domains.

Taken together, these directions underscore that routing is not merely a heuristic but a general design principle for collective intelligence systems. By explicitly deciding who speaks, what they report, and when silence is preferable, future work can move beyond one-size-fits-all aggregation rules toward adaptive, context-sensitive collective decision making.